\documentclass[11pt]{article}
\usepackage{amssymb,cite}
\def\ben{\begin{equation}}
\def\een{\end{equation}}

  \let\n=\nu

\let\C=\Chi

\def\nn{\nonumber} \def\bd{\begin{document}} \def\ed{\end{document}}
\def\ds{\documentstyle} \let\fr=\frac \let\bl=\bigl \let\br=\bigr
\let\Br=\Bigr \let\Bl=\Bigl
\let\bm=\bibitem
\let\na=\nabla
\let\pa=\partial \let\ov=\overline
\newcommand{\be}{\begin{equation}}
\newcommand{\ee}{\end{equation}}
\def\ba{\begin{array}}
\def\ea{\end{array}}
\def\ft#1#2{{\textstyle{{\scriptstyle #1}\over {\scriptstyle #2}}}}
\def\fft#1#2{{#1 \over #2}}
\def\del{\partial}
\def\vp{\varphi}
\def\sst#1{{\scriptscriptstyle #1}}
\def\oneone{\rlap 1\mkern4mu{\rm l}}
\def\td{\tilde}
\def\wtd{\widetilde}
\def\ie{\rm i.e.\ }
\def\dalemb#1#2{{\vbox{\hrule height .#2pt
        \hbox{\vrule width.#2pt height#1pt \kern#1pt
                \vrule width.#2pt}
        \hrule height.#2pt}}}
\def\square{\mathord{\dalemb{6.8}{7}\hbox{\hskip1pt}}}
\newcommand{\ho}[1]{$\, ^{#1}$}
\newcommand{\hoch}[1]{$\, ^{#1}$}
\newcommand{\bea}{\begin{eqnarray}}
\newcommand{\eea}{\end{eqnarray}}
\newcommand{\ra}{\rightarrow}
\newcommand{\lra}{\longrightarrow}
\newcommand{\Lra}{\Leftrightarrow}
\newcommand{\ap}{\alpha^\prime}
\newcommand{\bp}{\tilde \beta^\prime}
\newcommand{\tr}{{\rm tr} }
\newcommand{\Tr}{{\rm Tr} }
\def\0{{\sst{(0)}}}
\def\1{{\sst{(1)}}}
\def\2{{\sst{(2)}}}
\def\3{{\sst{(3)}}}
\def\4{{\sst{(4)}}}
\def\5{{\sst{(5)}}}
\def\6{{\sst{(6)}}}
\def\7{{\sst{(7)}}}
\def\8{{\sst{(8)}}}
\def\n{{\sst{(n)}}}
\def\cA{{{\cal A}}}
\def\cB{{{\cal B}}}
\def\cF{{{\cal F}}}
\def\cH{{{\cal H}}}
\def\tV{\widetilde V}
\def\tW{\widetilde W}
\def\tH{\widetilde H}
\def\tE{\widetilde E}
\def\tF{\widetilde F}
\def\tA{\widetilde A}
\def\im{{{\rm i}}}
\def\tY{{{\wtd Y}}}
\def\ep{{\epsilon}}
\def\vep{{\varepsilon}}
\def\R{\rlap{\rm I}\mkern3mu{\rm R}}
\def\bD{{{\bar D}}}
\def\R{\rlap{\rm I}\mkern3mu{\rm R}}
\def\bD{{{\bar D}}}
\def\R{{{\mathbb R}}}
\def\C{{{\mathbb C}}}
\def\H{{{\mathbb H}}}
\def\CP{{{\mathbb C}{\mathbb P}}}
\def\RP{{{\mathbb R}{\mathbb P}}}
\def\Z{{{\mathbb Z}}}
\def\bA{{{\mathbb A}}}
\def\bB{{{\mathbb B}}}
\def\bC{{{\mathbb C}}}
\def\bD{{{\mathbb D}}}
\def\bE{{{\mathbb E}}}
\def\bZ{{{\mathbb Z}}}
\def\Re{{{\mathfrak{Re}}}}
\def\Im{{{\mathfrak{Im}}}}
\def\cosec{{\,\hbox{cosec}\,}}
\def\Gm{{\Gamma_{\!\! -}}}
\def\Gp{{\Gamma_{\!\! +}}}
\def\stan{{standard }}
\def\nonstan{{supernumerary }}
\def\FF2{{ {}_{\sst 2}F_{\sst 1} }}
\def\FFF{{ {}_{\sst 3}F_{\sst 2} }}
\def\const{\rm constant}

\newcommand{\tamphys}{\it Center for Theoretical Physics,
Texas A\&M University, College Station, TX 77843}

\newcommand{\upenn}{\it Department of Physics and Astronomy,\\ University
of Pennsylvania, Philadelphia, PA 19104}

\newcommand{\damtp}{\it DAMTP, Centre for Mathematical Sciences,
 Cambridge University,\\  Wilberforce Road, Cambridge CB3 OWA, UK}

\newcommand{\brussels}{\it Physique Th\'eorique et Math\'ematique,
Universit\'e Libre de Bruxelles,\\ Campus Plaine C.P. 231, B-1050
Bruxelles, Belgium}

\newcommand{\alberta}{\it Theoretical Physics Institute, 412 Physics Lab., 
University of Alberta,\\ Edmonton, Alberta T6G 2J1, Canada{\hoch 3}}

\newcommand{\auth}{ G.W. Gibbons{\hoch *}, H. L\"u{\hoch {\ddagger 1}}, 
Don N. Page{\hoch {\ddagger,\dagger 2}}
and C.N. Pope\hoch{\ddagger 1}}

\begin{document}
\begin{flushright}

DAMTP-2004-47\ \ \
Alberta Thy 09-04\ \ \
MIFP-04-09\\
{\bf hep-th/0409155}\\
May\  2004
\end{flushright}

\vspace{10pt}

\begin{center}

{\large {\bf Rotating Black Holes in Higher Dimensions \\
              with a Cosmological Constant }}

\vspace{20pt}
\auth

\vspace{20pt}

\hoch{*}\damtp

\vspace{10pt}

\hoch{\ddagger}{\it George P. \& Cynthia W. Mitchell
Institute for Fundamental Physics,\\ Texas A\& M University,
College Station, TX 77843-4242, USA}

\vspace{10pt}

\hoch{\dagger}\alberta


\vspace{40pt}

\underline{ABSTRACT}
\end{center}

We present the metric for a rotating black hole with a cosmological
constant and with arbitrary angular momenta in all higher dimensions.
The metric is given in both Kerr-Schild and Boyer-Lindquist form.  In
the Euclidean-signature case, we also obtain smooth compact Einstein
spaces on associated $S^{D-2}$ bundles over $S^2$, infinitely many for
each odd $D \ge 5$.  Applications to string theory and M-theory are
indicated.

{\vfill\leftline{}\vfill \vskip 10pt \footnoterule 
{\footnotesize {\hoch 1}
Research supported in part by DOE grant
DE-FG03-95ER40917.}

{\footnotesize {\hoch 2} 
Research
  supported in part by the Natural Science and Engineering Research
  Council of Canada.}

{\footnotesize {\hoch 3}
Permanent address for D.N.P.}
}

\pagebreak

\newpage


  In recent years there has been a strong interest, in both physics and
mathematics, in higher dimensional solutions of Einstein's equations.
Black holes are among the most important exact solutions in general
relativity, and so solutions describing higher-dimensional black holes
are of particular significance.  The first general rotating black hole
solutions in higher dimensions were given by Myers and Perry
\cite{MyersPerry}, in the case that the cosmological constant
vanishes.  These have since been used extensively in string and
M-theory calculations.  More recently, interest has grown in Einstein
metrics with a cosmological constant, both cosmologically in four
dimensions, and in fundamental theories of nature in higher
dimensions.  In fact in four dimensions, Carter \cite{Carter} had
already found a generalization of the Kerr solution with a
cosmological constant and asymptotically de Sitter or anti-de Sitter
boundary conditions (the Kerr-de Sitter metric).  Hawking, Hunter and
Taylor-Robinson \cite{hawhuntay} generalized Carter's solution to five
dimensions, with arbitrary angular momenta, and to all dimensions with
just one nonzero angular momentum parameter.

    In a recent development, Tasinato {\it et al.} have shown that the
Kerr solutions in five dimensions or higher (with zero cosmological
constant) may be interpreted as time-dependent S-brane solutions of
string or M-theory \cite{tasi}.  (See also related work on twisted
S-branes, and their relation to Kerr solutions, in four dimensions
\cite{wang} and in higher dimensions \cite{lupor}.)  An important
question is how the results of this work on time-dependent
cosmological backgrounds in string and M-theory are affected by a
non-vanishing cosmological term.  This requires explicit solutions
generalising the higher-dimensional Kerr solution to the case when the
cosmological constant is nonzero.

Another area of string and M-theory where solutions with non-vanishing
cosmological constant are needed is in the AdS/CFT correspondence.
Following the pioneering work of \cite{hawhuntay} in five dimensions,
the principal remaining cases of interest are in dimensions six and
seven.  An important application of our new metrics is to study the
thermodynamics of rotating black holes in higher dimensional anti-de
Sitter backgrounds, especially those of relevance for the AdS/CFT
correspondence.  Recent work in \cite{gibperpop}, using our new
metrics, has provided complete results for the masses, and other
thermodynamic quantities, in all dimensions.  This has clarified
certain inconsistencies in previous literature, and in fact having the
results available in all dimensions has also helped to settle some
previous residual inconsistencies in four dimensions.

    A further striking application of the Kerr-de Sitter metrics is in
the Euclidean-signature regime, where in four dimensions they
provided, by analytic continuation, the first non-singular and compact
inhomogeneous Einstein metrics with positive-definite signature and
positive scalar curvature \cite{pagekerr}.  One application of this metric
is as an instanton mediating creation of the universe ``from nothing''
\cite{gibbons}.  This has been generalised in \cite{hassakyas} to 
five dimensions, producing an infinite family of non-singular
Einstein metrics which, for example, using the AdS/CFT correspondence,
provide infinitely many supersymmetry-breaking ground states for
${\cal N}=4$ supersymmetric Yang-Mills theory.  In a recent development, 
it has been shown that in limiting cases our new Kerr-de Sitter metrics
in the Euclidean regime give rise to infinite families of Einstein-Sasaki
metrics \cite{hassakyas2}, which can provide supersymmetric backgrounds
of importance for the AdS/CFT correspondence.

   Motivated by these considerations, we present here our basic
results for rotating black hole metrics in all higher dimensions with
a cosmological constant and with arbitrary angular momenta.  A more
detailed treatment is given in \cite{glpp}.

   Let the dimension of spacetime be $D = 2N + \ep + 1 \ge 4$, with $N
= [(D-1)/2]$ being the number of orthogonal spatial 2-planes, each of
which can have a rotation parameter $a_i$.  Thus $\ep=(D-1)$ mod 2.
Let $\phi_i$ be the $N$ azimuthal angles in the $N$ orthogonal
2-planes, each with period $2\pi$.  Let the remaining $N + \ep$
spatial dimensions be parameterized by a radial coordinate $r$ and by
$N + \ep$ `direction cosines' $\mu_i$ obeying the constraint
\be
\sum_{i=1}^{N+\ep} \mu_i^2 =1\,,
\label{constraint}
\ee
where $0\le \mu_i\le 1$ for $1\le i\le N$, and (for even $D$) 
$-1\le \mu_{N+1}\le 1$.

The Kerr-de Sitter metrics we have found satisfy $R_{\mu\nu} =
(D-1)\,\lambda\, g_{\mu\nu}$, and are given in Kerr-Schild form
\cite{KerrSchild} by
\be
ds^2 = d\bar s^2 + \fft{2M}{U}\, (k_\mu\, dx^\mu)^2\,,\label{kds}
\ee
where the de Sitter metric $d\bar s^2$, the null 1-form $k_\mu$, and
the function $U(r,\mu_i)$ are given by
\bea
d\bar s^2 &=& - W\,(1-\lambda \, r^2)\, 
dt^2 + F\, dr^2 + \sum_{i=1}^{N+\ep} \fft{r^2 + a_i^2}{1+\lambda\, a_i^2} 
\,\, d\mu_i^2 + \sum_{i=1}^N \fft{r^2 + a_i^2}{1+\lambda\, a_i^2}
\, \, \mu_i^2\, d\phi_i^2 \nn \\
&& \qquad + 
 \fft{\lambda}{W\, (1-\lambda\, r^2)}\, 
\Big( \sum_{i=1}^{N+\ep} 
\fft{(r^2 + a_i^2)\, \mu_i\, d\mu_i}{1+\lambda\, a_i^2}
   \Big)^2\,,\label{ds}\\ 
k_\mu\, dx^\mu &=& F\, dr + W\, dt - 
\sum_{i=1}^N \fft{a_i\, \mu_i^2}{1+\lambda\, a_i^2} \,  d\phi_i \,,
\label{case2}\\
U &=& r^{\ep}\, \sum_{i=1}^{N+\ep} \fft{\mu_i^2}{r^2 + a_i^2}\, 
\prod_{j=1}^N (r^2 + a_j^2)\,,\label{case3}
\eea
where the functions $W(\mu_i)$ and $F(r,\mu_i)$ are defined to be
\be
W \equiv \sum_{i=1}^{N+\ep} \fft{\mu_i^2}{1+\lambda\, a_i^2}\,,\qquad
F\equiv \fft{1}{1-\lambda\, r^2}\, \, 
  \sum_{i=1}^{N+\ep} \fft{r^2 \, \mu_i^2}{r^2+a_i^2}\,.\label{WFdef}
\ee

    We have been led to these metrics by putting the previously-known
$D=4$ and $D=5$ Kerr-de Sitter metrics into Kerr-Schild form, and
making natural generalisations to higher dimensions.  We have
explicitly checked that they obey the Einstein equation for all
physicially interesting cases $D \le 11$.  Since $D\le 11$ is not
distinguished in any way in the general expressions for the metrics,
we are confident that they are valid in all dimensions.  Furthermore,
if all rotations $a_i$ except one are set to zero, our expressions
reduce to those obtained in \cite{hawhuntay} in any dimension.  Finally,
we note that if the cosmological constant is set to zero, our metrics 
reduce to those found by Myers and Perry \cite{MyersPerry}.

   One may eliminate cross terms with $dr$ by passing to generalized
Boyer-Lindquist coordinates
\be
dt=d\tau + \fft{2M\, dr}{(1 - \lambda\, r^2)(V-2M)}\,,\qquad
d\phi_i = d\varphi_i
+ \fft{2M\,a_i\, dr}{(r^2 + a_i^2)(V-2M)}\,.\label{coordtrans}
\ee
The Kerr-de Sitter metrics then have the form
\bea
ds^2 &=& - W\, (1 -\lambda\, r^2)\, d\tau^2
 + \fft{2M}{U}\Bigl(W\,d\tau
 - \sum_{i=1}^N \fft{a_i\, \mu_i^2\, d\varphi_i}
  {1 + \lambda\, a_i^2}\Bigr)^2
 + \sum_{i=1}^N \fft{r^2 + a_i^2}{1 + \lambda\, a_i^2}\,\mu_i^2\,
    d\varphi_i^2 \nn\\
&&
 + \fft{U\, dr^2}{V-2M}
 + \sum_{i=1}^{N+\ep} \fft{r^2 + a_i^2}{1 + \lambda\, a_i^2}\, d\mu_i^2
 + \fft{\lambda}{W\, (1-\lambda r^2)}
    \Bigl( \sum_{i=1}^{N+\ep} \fft{r^2 + a_i^2}{1 + \lambda\, a_i^2}
    \, \mu_i\, d\mu_i\Bigr)^2 \,,\label{bl}
\eea
where $V(r)$ is defined by
\be
V\equiv \fft{U}{F} \equiv r^{\ep-2}\, (1-\lambda\, r^2)\, 
   \prod_{i=1}^N (r^2 + a_i^2)\, .
\ee

The Kerr-de Sitter metrics have Killing horizons at $r=r_{\sst H}$,
where $V(r_{\sst H}) = 2M$ and where the Killing vector field
\be
l = \fft{\del}{\del t} + 
\sum_{i=1}^N 
\fft{a_i\, (1-\lambda\, r_{\sst H}^2)}{r_{\sst H}^2+a_i^2}\, 
\fft{\del}{\del \phi_i}\,
= \, \fft{\del}{\del \tau} + 
\sum_{i=1}^N 
\fft{a_i\, (1-\lambda\, r_{\sst H}^2)}{r_{\sst H}^2+a_i^2}\, 
\fft{\del}{\del \varphi_i}\,
\ee
coincides with the null generator of the horizon.
The Kerr-Schild coordinates extend through the future
horizon.  By contrast, the Boyer-Lindquist coordinates are valid
either outside the horizon or inside the horizon.  It is the latter 
case, in which $r$ plays the r\^ole of the time coordinate, that
is relevant for time-dependent S-brane solutions.

   On the horizon, the Killing vector $l$ obeys $l^\mu\, \nabla_\mu\, 
l_\nu = \kappa\, l_\nu$, 
where the surface gravity, constant on each connected
component of the horizon, is given by
\be
\kappa= r_{\sst H}(1-\lambda\, r_{\sst H}^2)\,\left(\sum_{i=1}^{N} 
\fft1{r_{\sst H}^2 + a_i^2}+\fft{\ep}{2r_{\sst H}^2}\right)
 - \fft1{r_{\sst H}}\,,\label{kappa}
\ee
The area of the horizon is
given by
\be
A_{\sst H} = {\cal A}_{D-2}\, r_{\sst H}^{\ep-1}\prod_{i=1}^{N} 
   \fft{r_{\sst H}^2 + a_i^2}{1+\lambda\, a_i^2}\,,
\ee
where
\be
{\cal A}_{m} = \fft{2 \pi^{(m+1)/2}}{
   \Gamma[ (m +1)/2]}\label{snvol}
\ee
is the volume of the unit $m$-sphere.

    We can pass from the Lorentzian-signature Kerr-de Sitter metrics
(\ref{bl}) to Euclidean-signature Einstein metrics by making the
Boyer-Lindquist time coordinate $\tau$ and the rotation parameters
$a_i$ all purely imaginary.  The generic local Einstein metrics do not
give smooth complete compact Einstein spaces, but for $\lambda > 0$,
we can choose discrete special values for $a_i$ and $M$, as in four
dimensions in \cite{pagekerr} and in five dimensions in
\cite{hassakyas}, to get complete non-singular metrics.

   The idea is that $i\tau$ becomes an angular coordinate with the
appropriate period required to avoid a conical singularity at one root
of $V(r)-2M$, say at $r=r_1$.  We call this the black hole horizon, by
analogy with the Lorentzian-signature case.  If $r$ ranges from $r_1$
to a second root of $V(r)-2M$, say at $r=r_2$ (which we shall call the
cosmological horizon), we require the same period of $i\tau$ to avoid
a conical singularity at $r=r_2$.  Thus the surface gravities at $r_1$
and at $r_2$ must be identical, which can be accomplished by choosing
$M$ so that $r_1$ approaches $r_2$.  In this limit $g_{rr}$ diverges
in just the right way that the proper distance between the two roots
or horizons approaches a nonzero finite limit.  In the limiting
process, the period of $i\tau$ goes to infinity, but the metric length
of its orbit remains finite.  After appropriately rescaling $r$ and
$i\tau$, one arrives at a finite metric.

   The remaining conditions for regularity are that in each 2-plane 
with a nonzero rotation parameter
$a_i$, the black hole horizon rotate an integer number
$k_i$ times, relative to the cosmological horizon, during
one period of the Euclidean time coordinate $i\tau$.  More details 
are given in \cite{glpp}.  These conditions place $N$ constraints on the 
$N$ rotation parameters $a_i$.

   One obtains \cite{glpp} smooth compact
Einstein metrics of the form
\bea
\lambda ds^2 &&\!\!\!\!\!\!\!\!\!
 = \fft{(1\!+\!A)z(\mu_i)}{4A\!+\!2A^2\!+\!2B}\, 
   (d\chi^2\! +\! \sin^2\chi\, d\psi^2) +  
\sum_{i=1}^{N+\ep}\, \fft{(1\!+\!A)d\mu_i^2}{1\!+\!A\!+\!x_i}\,
  + \fft{A}{w(\mu_i)}\, \Big( \sum_{i=1}^{N+\ep}
 \fft{(1\!+\!A)\mu_i\, d\mu_i}{1\!+\!A\!+\!x_i} \Big)^2\, \nn \\
\ +&&\!\!\!\!\!\!\!\!\!\!
\sum_{i=1}^N \fft{(1\!+\!A)\mu_i^2}{1\!+\!A\!+\!x_i}\, 
    (d\varphi_i\! +\! k_i\, \sin^2\fft{\chi}{2}\, d\psi)^2
 - \fft{1\!+\!A}{z(\mu_i)}\, \Big[\sum_{i=1}^N \fft{\sqrt{x_i\!+x_i^2}\,
  \mu_i^2}
  {1\!+\!A\!+\!x_i}\,
    (d\varphi_i\! +\! k_i\, \sin^2\fft{\chi}{2}\, d\psi)\Big]^2 \,,
\nn \\
\label{eumetric}
\eea 
where
\bea
&& A \equiv \sum_{i=1}^N x_i - \fft{\ep}{2}\,,\qquad
B \equiv \sum_{i=1}^N x_i^2 + \fft{\ep}{2}\, , \nn \\
&&w(\mu_i) \equiv \sum_{i=1}^{N+\ep}\fft{A x_i \mu_i^2}{1+A+x_i} \,, \qquad
z(\mu_i) \equiv \sum_{i=1}^{N+\ep} x_i \mu_i^2 \, ,
\label{ABCdef}
\eea
and where the parameters $x_i$ must be chosen so that each
\be
k_i = \fft{2(1+A+x_i) \sqrt{x_i+x_i^2} }{2A+A^2+B}
\label{reg3}
\ee
is an integer.

In terms of the constant parameters $x_i$ defined here and the
resulting auxiliary constants $A$ and $B$, the roots of $V(r)-2M$ and
the parameters of the Kerr-de Sitter metric (\ref{bl}) are given by
\be
r_1 = r_2 = \sqrt{\fft{1+A}{\lambda A}} \,,\qquad
M = -\fft{r_1^{D-3}}{2A}\, \prod_{i=1}^N (-x_i^{-1})\,,\qquad
a_i = \im\, r_1\, \sqrt{\fft{1+x_i}{x_i}}\,.
\ee

One can see that for a real Euclidean metric, one needs either all
$x_i \le -1$ or all $x_i > 0$.  In the former case, we showed
\cite{glpp} that all but one $x_i$ must be $-1$, corresponding to only
one $a_i$ and $k_i$ nonzero, and that the only nonzero $k_i$ allowed
is $k_i=1$.  The resulting solutions were first given in
\cite{pagekerr} for $D=4$ and in \cite{hassakyas} for higher $D$.  In
the latter case, where all $x_i$ are positive, which is allowed only
for odd $D$, we showed \cite{glpp} that all possible sets of purely
positive integers $k_i$ lead to unique solutions for $x_i > 0$ and to
unique regular compact Einstein metrics, though $k_1=k_2=1$ for $D=5$
leads to $x_1=x_2=\infty$ rather than to finite solutions for the
$x_i$.  This and certain other cases in which one or more $x_i=\infty$
also lead to regular metrics \cite{hassakyas}.

Except for the $D=5$ solutions with $k_1 \ge 1$ and $k_2 \ge 1$, which
were given by \cite{hassakyas}, our compact Einstein metrics with all
$k_i > 0$ in odd $D$ are new.  Because all sets of positive $k_i$ are
allowed, for all odd $D\ge 5$ we get an infinite set of smooth compact
Einstein metrics on $S^2 \times S^{D-2}$ when $\sum_i k_i$ is even,
and an infinite set of smooth compact Einstein metrics on the
nontrivial $S^{D-2}$ bundle over $S^2$ when $\sum_i k_i$ is odd.

\bigskip
\noindent{{\bf Acknowledgements}}:  We thank Tekin Dereli, 
Sean Hartnoll, and
Yukinori Yasui for helpful discussions.  G.W.G. and D.N.P. are
grateful to the George P. \& Cynthia W. Mitchell Institute for
Fundamental Physics for hospitality during the course of this work.

\end{document}